# A Search for New Galactic Microquasars

G.S. Tsarevsky (1,2), E.P. Pavlenko (3), R.A. Stathakis (4),
N.S. Kardashev (2), O.B. Slee (1)

(1) Australia Telescope National Facility, Sydney,
(2) Astro Space Centre, Moscow, Russia,
(3) Crimean Astrophysical Observatory, Ukraine,
(4) Anglo-Australian Observatory, Sydney

**Abstract**

A complete sample of bright ROSAT sources with hard XRB-like spectra in the Galactic Plane ( $|b| < 15^o$ ) has been tentatively identified with radio sources in the GB6/PMN/NVSS surveys, and subsequently observed with the Australia Telescope Compact Array and the Very Large Array. Most of them are unresolved at the sub-arcsec scale and have flat or inverted spectra.

Precise radio coordinates have made unambiguous optical identifications possible, which, after the removal of galaxies, yielded a final list of 40 microquasar candidates. They are successfully going through the moderate dispersion spectroscopy by the 4-m telescope of the Anglo-Australian Observatory.

Our goal is to obtain evidence for a characteristic accreting behaviour and establish binarity, thence permitting actual microquasar classification. VLBI observations of the brightest candidates are also underway.

We expect some of these objects could be QSOs, or radio galaxies, or cataclysmic variables. However, this would be a valuable by-product of the proposed program.

Photometry of these objects dedicated to find possible eclipses and, also, characteristic accreting disc driven flares.

## 1. Introduction: Population of Microquasars in our Galaxy

Accretion onto a supermassive black hole with a strong surrounding magnetic field can supply the necessary energy for AGNs (Kardashev, 1995). Inside our own galaxy, accretion from a stellar component onto a black hole (or neutron star) in a close binary system can produce a similar kind of phenomenon.

X-ray observations made by UHURU in 1978 attracted attention to the peculiar object SS 433 located in the very centre of the supernova remnant W50.
When the orbital period was first determined, Shklovski (1978) suggested that SS 433 is a binary system associated with the ejection of relativistic particles, which are responsible for the strong, periodic radio emission.

Many observations of SS 433 led to the conclusion that the system is a close binary consisting of a massive OB star and a neutron star or a black hole surrounded by a bright accretion disk opaque to X-rays. SS 433 and similar objects have been assigned to a special class called "microquasars" (see comprehensive review by Mirabel and Rodriguez, 1999).

Only about 30 of ~280 known X-ray binaries (XRBs) have been detected in radio (Fender et al. 1997), and only a few of them have characteristic radio emission and morphology associated with the microquasars' family. Radio images of such objects bear a striking similarity to the structures of AGN: they have a compact core and two-sided jets of relativistic particles. Flux variability and super-luminal motions are also quite common for the microquasars. GRO J1655-40 is a representative object of this class (Tingay at al. 1995). First discovered in
X-rays, it produces relativistic radio jets with $\beta = 0.92$, and has an angular extent of 1 arcsec. It is intention of the project described here to search for similar features with the aim of increasing the number of known microquasars.

## 2. A Search for new Microquasars: Selection Criteria and Sample Description

First, we have selected sources from the ROSAT All-Sky Survey Bright Source Catalogue (Voges et al. 1999) by the following criteria:
 1) close to the Galactic Plane ($|b| < 20^o$);
 2) with extent less than 10 arcsec, i.e. point-like sources; and
 3) having X-ray spectra typical of the XRBs.
The last selection has been undertaken in accordance with criteria developed by Motch et al. (1998). They found that XRBs in a 'two-colour' HR1-HR2 diagram show a strong concentration in the narrow strip restricted to the following values:
$$0.90 < HR1 \leq 1.00;$$
and
$$0.25 < HR2 \leq 1.00$$
A total of 243 1RXS sources were found to fit these criteria.
The X-ray sources with large errors in HR1 and HR2 were discarded from the list. We also discarded X-ray sources with position errors of more than 15 arcsec.
Thus a total of 243 1RXS sources were found to fit these criteria.
Secondly, we have identified the sources in this basic sample with radio sources in the PMN, GB6 and NVSS catalogues. To increase the probability of radio identification, we applied strong criteria for coordinate coincidence, within the 1-sigma errors of the catalogues. After rejection of some already known objects of various types and, also, radio sources weaker than 5 mJy, this secondary list consists of 48 sources.

All these have been observed with the Australia Telescope Compact Array and the Very Large Array. Most of them were successfully detected, and consequently unambiguously optically identified. They are intended for a follow-up spectro-scopic and photometric study to establish their binary nature. We have already started a CCD monitoring of northern objects in the Crimean Astrophysical Observatory.

Some of our objects have already been observed in the Anglo-Australian Observatory 4-m telescope service time using the RGO low resolution spectrograph. Source j1628-41a has shown a signature of a K5 star with strong and variable H$\alpha$ emission.

Such emission could be closely connected with accretion process in a binary system as described above. Therefore this object deserves an intensive search by conventional spectroscopy and photometry, to find an evidence of its duplicity and accretion driven flares. Together with its hard X-ray spectrum and possible jet-like radio emission is stand like our best microquasar candidate.